\documentclass[sigconf, nonacm]{acmart}

\usepackage{listings}
\usepackage[inkscapeopt={-D}]{svg}
\usepackage{multirow}
\usepackage{pifont}
\usepackage{rotating}
\usepackage{xcolor}

\lstdefinestyle{yaml}{
  basicstyle=\ttfamily\small,
  sensitive=false,
  comment=[l]{\#},
  commentstyle=\color{gray},
  showstringspaces=false,
  emphstyle=\bfseries,
  emph={get_method_code,description,method_signature,type,class_name,default},
  identifierstyle=,
  keywords={null, string},
}

\definecolor{javared}{rgb}{0.6,0,0}          
\definecolor{javagreen}{rgb}{0.25,0.5,0.35}  
\definecolor{javapurple}{rgb}{0.5,0,0.35}    
\definecolor{javadocblue}{rgb}{0.25,0.35,0.75} 

\copyrightyear{2026}
\acmYear{2026}
\setcopyright{cc}
\setcctype{by}
\def\axtext{\footnotesize\sffamily}
\def\ticktext{\scriptsize\sffamily}

\begin{document}

\title{Coverage-Guided Multi-Agent Harness Generation for Java Library Fuzzing}

\author{Nils Loose}
\email{n.loose@uni-luebeck.de}
\orcid{0009-0003-6243-1623}
\affiliation{%
  \institution{University of Lübeck}
  \department{Institute for IT Security}
  \city{Lübeck}
  \country{Germany}
}
\author{Nico Winkel}
\email{nico.winkel@student.uni-luebeck.de}
\orcid{0009-0008-8538-6892}
\affiliation{%
  \institution{University of Lübeck}
  \department{Institute for IT Security}
  \city{Lübeck}
  \country{Germany}
}
\author{Kristoffer Hempel}
\email{k.hempel@uni-luebeck.de}
\orcid{0009-0008-9159-4268}
\affiliation{%
  \institution{University of Lübeck}
  \department{Institute for IT Security}
  \city{Lübeck}
  \country{Germany}
}
\author{Felix Mächtle}
\email{f.maechtle@uni-luebeck.de}
\orcid{0009-0009-2431-0322}
\affiliation{%
  \institution{University of Lübeck}
  \department{Institute for IT Security}
  \city{Lübeck}
  \country{Germany}
}
\author{Julian Hans}
\email{julian.hans@student.uni-luebeck.de}
\orcid{0000-0003-0763-3241}
\affiliation{%
  \institution{University of Lübeck}
  \department{Institute for IT Security}
  \city{Lübeck}
  \country{Germany}
}
\author{Thomas Eisenbarth}
\email{thomas.eisenbarth@uni-luebeck.de}
\orcid{0000-0003-1116-6973}
\affiliation{%
  \institution{University of Lübeck}
  \department{Institute for IT Security}
  \city{Lübeck}
  \country{Germany}
}

\renewcommand{\shortauthors}{Loose, et al.}

\begin{abstract}
Coverage-guided fuzzing has proven effective for software testing, but targeting library code requires specialized fuzz harnesses that translate fuzzer-generated inputs into valid API invocations. Manual harness creation is time-consuming and requires deep understanding of API semantics, initialization sequences, and exception handling contracts.
We present a multi-agent architecture that automates fuzz harness generation for Java libraries through specialized LLM-powered agents. Five ReAct agents decompose the workflow into research, synthesis, compilation repair, coverage analysis, and refinement. Rather than preprocessing entire codebases, agents query documentation, source code, and callgraph information on demand through the Model Context Protocol, maintaining focused context while exploring complex dependencies.
To enable effective refinement, we introduce method-targeted coverage that tracks coverage only during target method execution to isolate target behavior, and agent-guided termination that examines uncovered source code to distinguish productive refinement opportunities from diminishing returns.
We evaluated our approach on seven target methods from six widely-deployed Java libraries totaling 115,000+ Maven dependents. Our generated harnesses achieve a median 26\% improvement over OSS-Fuzz baselines and outperform Jazzer AutoFuzz by 5\% in package-scope coverage. Generation costs average \$3.20 and 10 minutes per harness, making the approach practical for continuous fuzzing workflows. During a 12-hour fuzzing campaign, our generated harnesses discovered 3 bugs in projects that are already integrated into OSS-Fuzz, demonstrating the effectiveness of the generated harnesses.
\end{abstract}



\maketitle

\section{Introduction}
\begin{figure*}[t!]
    \centering
    \includesvg[width=.8\linewidth]{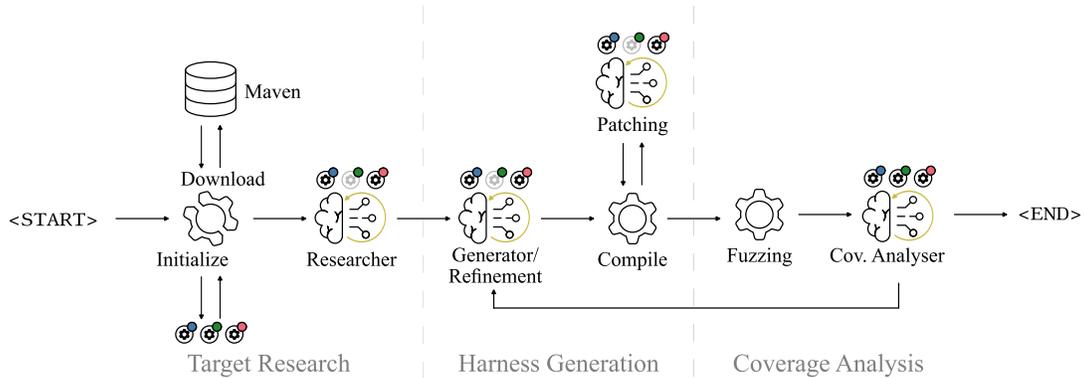}
    \caption{Schematic overview of the harness generation workflow. Agents are ReAct agents with specialized tool access.}
    \Description{Schematic overview of the harness generation workflow. Agents are ReAct agents with specialized tool access. The Research Agent gathers information about the target method using the javadoc MCP and code MCP. The generation agent generates an initial harness using the gathered information with additional query capability. The patching agent attempts to repair the harness if compilation fails, using compiler error messages and the code and javadoc MCP to fix any issues. The coverage analysis agent runs after fuzzing to analyze method-targeted coverage results using the callgraph MCP. Finally, the refinement agent uses coverage reports and uncovered code locations to iteratively improve the harness.}%
    \label{fig:overview}
\end{figure*}

Coverage-guided fuzzing has become a fundamental technique for discovering bugs and vulnerabilities in software systems. When applied to library code, its effectiveness depends on the availability of high-quality fuzz harnesses. A harness serves as an adapter between the fuzzer and the target library, transforming unstructured byte sequences into valid API invocations that exercise library logic. Manual harness creation remains a significant obstacle to widespread fuzzing adoption. Developers must understand API contracts, construct valid object states, synthesize realistic call sequences, and implement appropriate exception handling. This time-intensive process limits the number of library APIs that receive comprehensive fuzzing coverage. This challenge is particularly acute for Java libraries, which remain underrepresented in continuous fuzzing infrastructure and research, despite the widespread deployment of Java applications in production systems~\cite{oss-fuzz-gen}. 

Existing automated harness generation approaches face distinct and complementary limitations. Usage-based methods mine API interaction patterns from consumer code~\cite{DBLP:conf/sigsoft/BabicBCIKKLSW19:FUDGE,DBLP:conf/sp/JeongJYMKJKSH23:UTopia} but require access to substantial client corpora that may be unavailable for specialized or newly released libraries. Structure-based approaches derive harnesses from type signatures and interface specifications~\cite{DBLP:conf/icse/GreenA22:GraphFuzz,DBLP:conf/icse/ShermanN25:OGHarn} but struggle with implicit preconditions and often rely on domain-specific heuristics that limit generalizability. Feedback-driven methods employ iterative refinement based on runtime signals~\cite{DBLP:conf/uss/ZhangLZZZZXLL0H23:Rubick} but typically apply fixed termination thresholds without semantic interpretation of coverage gaps. Recent LLM-based systems demonstrate progress through coverage-guided prompt evolution~\cite{DBLP:conf/ccs/LyuXCC24:PromptFuzz} and knowledge graph augmentation~\cite{DBLP:conf/icse/XuMZZCHLW25:CKGFuzzer}. However, these approaches lack mechanisms for iterative, query-driven exploration during generation. 

Large language models (LLMs) present new opportunities for automated harness generation by combining code synthesis capabilities with domain specific knowledge. However, directly applying LLMs to this task introduces distinct challenges. Preprocessing entire API surfaces exhausts available context windows, particularly for large library ecosystems with extensive dependency graphs. One-shot generation fails on complex libraries requiring multi-step initialization sequences or non-obvious preconditions. Coverage-based refinement risks semantic drift when agents lack mechanisms to interpret what the coverage gaps represent and whether they indicate addressable deficiencies or fundamental limitations. These observations suggest that effective harness generation requires an approach that integrates LLM reasoning with targeted program analysis, retrieves information on demand rather than preprocessing entire codebases, and interprets coverage feedback semantically rather than applying fixed numerical thresholds.

We address these challenges through a multi-agent architecture that decomposes harness generation into specialized reasoning tasks. Five ReAct~\cite{DBLP:conf/iclr/YaoZYDSN023:ReAct} agents handle distinct phases of the workflow. The research agent explores API documentation and source code to understand target method semantics. The synthesis agent transforms this understanding into initial harness implementations. The compilation agent diagnoses and repairs build errors through iterative refinement. The coverage analysis agent interprets coverage gaps by examining uncovered source code to determine whether further refinement is worthwhile. The refinement agent modifies harnesses to address identified coverage deficiencies. Rather than preprocessing entire API surfaces into static knowledge graphs, agents query information on demand through the Model Context Protocol~\cite{mcp} (MCP). This query-driven approach retrieves documentation for specific methods, source code for particular classes, and callgraph fragments rooted at selected invocations. The design maintains a focused context while exploring large dependency graphs, enabling autonomous handling of complex build configurations and iterative compilation repair.

Two technical mechanisms enable effective coverage-guided refinement. First, we introduce method-targeted coverage instrumentation that activates coverage tracking~\cite{jacoco} only during target method execution. Standard instrumentation measures all executed code, creating misaligned incentives where harnesses invoke unrelated utility methods to inflate coverage metrics. Our approach ensures that coverage measurements reflect target behavior rather than incidental framework initialization. Second, we implement agent-guided termination that interprets coverage gaps through source code analysis. The coverage analysis agent examines uncovered methods to distinguish addressable deficiencies such as missing input variants or unexplored API paths from fundamental limitations. This interpretation enables the system to stop refinement when it yields diminishing returns while continuing when concrete improvement strategies exist.

We evaluate our approach on seven target methods from six widely deployed Java libraries spanning parsers, JSON processors, and core utilities. The selected libraries total over 115,000 Maven dependents and represent real-world fuzzing targets. All targets have existing harnesses in OSS-Fuzz~\cite{oss-fuzz}, providing strong baselines for comparison. Our generated harnesses achieve a median improvement of 26\% in method-targeted coverage over OSS-Fuzz baselines and outperform both OSS-Fuzz and Jazzer AutoFuzz~\cite{jazzer} by 6\% and 5\% respectively under full package-scope coverage. Generation costs average \$3.20 and approximately 10 minutes per harness, demonstrating practical feasibility for integration into continuous fuzzing workflows. During 12-hour fuzzing campaigns with our harnesses, we  discovered 3 previously unreported bugs. These discoveries occurred in methods already covered by existing OSS-Fuzz harnesses, underlining the harnesses effectiveness.
\par
In summary, this work makes the following contributions.
\begin{itemize}
    \item A multi-agent architecture that integrates LLM reasoning with program analysis to automate harness generation for Java library APIs without requiring consumer code corpora or manual intervention.

    \item A query-driven tool interface using the Model Context Protocol that retrieves precisely scoped and preanalysed program information on demand, preventing context saturation while enabling exploration of large codebases.

    \item Method-targeted coverage information with agent-guided termination that interprets coverage gaps through source code analysis rather than applying fixed thresholds.

    \item Empirical validation on widely deployed libraries demonstrating competitive coverage with manually written baselines, practical generation costs, and discovery of three previously unreported bugs.
\end{itemize}

\begin{figure}[b]
    \centering
    \includesvg[width=.9\linewidth]{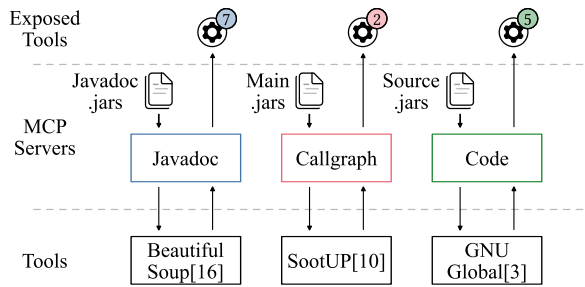}
    \caption{Overview of exposed tools through the model context protocol (MCP).}
    \Description{Overview of exposed tools through the model context protocol (MCP). Three MCP Servers expose documentation search (Javadoc MCP) using BeautifulSoup, call graph analysis (Call Graph MCP) using SootUp, and source code search (Code MCP) using GNU Global. Each MCP exposes a list of tools that can be queried by agents to retrieve relevant information on demand.}%
    \label{fig:mcp-design}
\end{figure}

\section{Preliminaries}

\paragraph{Harness Design for Coverage-Guided Fuzzing}
Coverage-guided fuzzers generate test inputs by mutating input sequences based on code coverage feedback. When fuzzing library APIs, a \emph{fuzz harness} serves as the entry point that translates fuzzer-generated bytes into valid API invocations. An effective harness must parse input bytes into appropriate data types, construct required object states to satisfy API preconditions, invoke target methods with derived arguments, and handle exceptions appropriately to distinguish expected error conditions from genuine bugs. The harness design directly impacts fuzzing effectiveness. A harness that exercises diverse API paths and satisfies complex preconditions enables the fuzzer to reach deeper code and discover latent bugs.

\paragraph{Tool-Augmented Reasoning}
Tool-augmented reasoning equips LLMs with external capabilities they can invoke during generation~\cite{DBLP:conf/nips/SchickDDRLHZCS23:Toolformer}. The ReAct (Reasoning and Acting) paradigm~\cite{DBLP:conf/iclr/YaoZYDSN023:ReAct} formalizes tool use as an interleaved process where the model alternates between reasoning steps that generate natural language explanations of its strategy and action steps that invoke tools and observe outputs. For code generation, tools typically include documentation search, source code retrieval, compilation, and test execution. Rather than providing all information upfront in a single prompt, ReAct agents query information on demand as their reasoning progresses.

\paragraph{Multi-Agent LLM Architectures}
Recent work has demonstrated that LLMs can be organized into multi-agent systems where distinct instances specialize in complementary subtasks~\cite{DBLP:journals/corr/abs-2312-13010:AgentCoder}. Rather than relying on a single monolithic prompt, multi-agent architectures decompose complex objectives into stages handled by specialized agents that communicate through structured message passing or a shared state. This decomposition enables separation of concerns which has been shown to improve both generation quality and success rates on challenging software benchmarks~\cite{DBLP:conf/iclr/0036YZXLL0DMYZ024:AgentBench}.

\paragraph{Model Context Protocol}
The Model Context Protocol~\cite{mcp} standardizes how LLM agents access external resources through structured tool interfaces. MCP defines a client-server architecture where agents act as clients that invoke tools exposed by servers managing data sources. Each tool accepts structured parameters and returns formatted responses optimized for LLM consumption. MCP enables query-driven information retrieval where agents request precisely scoped information as needed.

\section{Agent-Based Harness Generation}%
\label{sec:approach}
We present an agent-based approach to automated fuzzing harness generation that addresses several fundamental challenges of generating effective harnesses for complex library APIs. Our approach combines LLM-powered agents with static analysis and dynamic coverage feedback to iteratively construct and refine harnesses that achieve deep code coverage.
Figure~\ref{fig:overview} illustrates our workflow as a sequence of transformations that progressively refine a fuzzing harness. After initializing the environment by downloading library artifacts and preparing analysis infrastructure, the workflow proceeds through three phases: target research to understand API semantics, harness construction through code generation and compilation, and iterative coverage-guided refinement. Specialized ReAct agents~\cite{DBLP:conf/iclr/YaoZYDSN023:ReAct} orchestrate each phase, querying documentation and source code on demand to maintain focused context while exploring large dependency graphs.
\begin{lstlisting}[float=h,language=Java,caption={Harness showing selective coverage instrumentation through runtime control of JaCoCo's recording state.},label={lst:coverage-tracking},float=htbp,basicstyle=\small\ttfamily,numbers=left,numberstyle=\tiny\color{gray},frame=single]
public static void fuzzerTestOneInput(
      FuzzedDataProvider data) {
   RT.getAgent().setRecording(false); 
   // Parameter/ Instance preparation
   Options o = prepareOptions(data);
   try{
      RT.getAgent().setRecording(true);
      Parser.parse(o);
      RT.getAgent().setRecording(false);
   } catch (IllegalArgumentException var15) {
        RT.getAgent().setRecording(false);
        // Expected exception 
      }
}
\end{lstlisting}
\subsection{Static Analysis and Instrumentation}%
\label{subsec:static-analysis}
Our approach requires three preprocessing artifacts that enable efficient agent exploration and accurate coverage measurement. First, we extract API documentation from Javadoc HTML archives distributed with Maven artifacts, parsing method signatures and parameter descriptions using Beautiful Soup~\cite{beautifulsoup} to provide agents with concise API contracts indexed for query-based retrieval. Second, we index the source code using GTAGs~\cite{gnuglobal}, enabling efficient symbol resolution and context retrieval during agent reasoning. Lastly, we compute a static callgraph rooted at the target method using SootUp's Class Hierarchy Analysis~\cite{DBLP:conf/tacas/KarakayaSKBSLH24:SootUp}, traversing method invocations to depth~10 (depth~5 for large libraries). Each node records the method signature, enclosing class, and distance from the target, serving to scope coverage analysis to reachable methods and provide agents with structural context about call dependencies. Additionally, we implement method-targeted coverage instrumentation. Standard coverage instrumentation measures all executed code, creating a misaligned incentive for agents to invoke unrelated utility methods. We address this by extending JaCoCo~\cite{jacoco} with runtime toggling of coverage tracking, accessible through the runtime API. Using ASM~\cite{asm} for offline bytecode instrumentation, we wrap target method invocations to enable coverage recording only during target execution (Listing~\ref{lst:coverage-tracking}), ensuring metrics reflect the target's behavior rather than incidental framework initialization.

\begin{figure}[thb]
\centering
\includesvg[width=.9\linewidth]{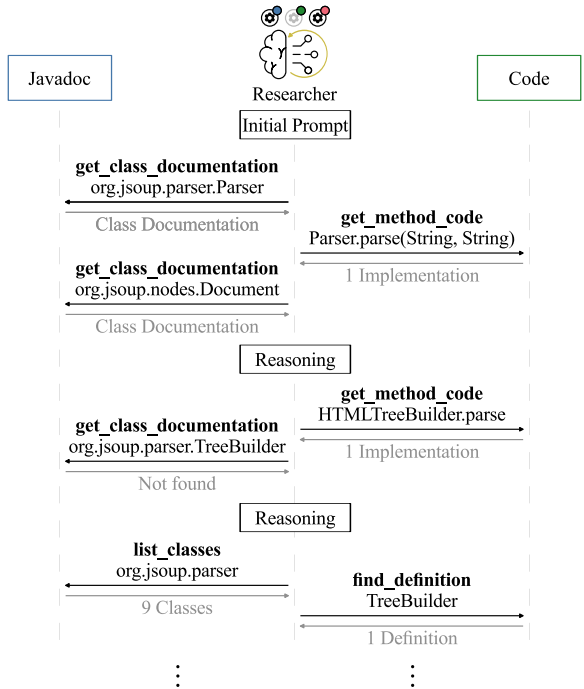}
\caption{Sequence diagram showing initial tool interactions during the research phase for \texttt{Jsoup.parse(String)}.}
\Description{Sequence diagram showing initial tool interactions during the research phase for \texttt{Jsoup.parse(String)}. The research agent iteratively requests information for the relevant classes and requests specific method implementations from the relevant MCP servers throughout the REPL loop.}%
\label{fig:research-sequence}
\end{figure}
\subsection{Tool-Augmented Exploration}%
\label{subsec:tool-augmented-exploration}
We expose preprocessing artifacts to agents through a query-based interface using the Model Context Protocol (MCP)~\cite{mcp}, enabling on-demand retrieval of tailored information as reasoning progresses.
We provide three tool categories: documentation queries, source code retrieval, and callgraph queries. Figure~\ref{fig:mcp-design} illustrates the MCP initialization process. Each tool accepts structured parameters (e.g., class name, method signature) and returns responses optimized for LLM consumption: concise method signatures, minimal code snippets, and depth-limited callgraph fragments.
To prevent exploration drift, we restrict tool access based on agent role. Table~\ref{tab:tool-availability} summarizes tool availability across the five ReAct agents. All agents have access to documentation and source code tools for foundational API understanding, while callgraph tools are restricted to the coverage analysis agent for interpreting coverage gaps. This role-based access control prevents agents from pursuing information irrelevant to their current task.

\begin{table}[b]
\caption{Tool availability across agents. MCP tools are provided by three Model Context Protocol servers.}
\centering
\setlength{\tabcolsep}{3pt}
\renewcommand{\arraystretch}{1.1}
\small
\begin{tabular}{c l | c c c c c}
\toprule
& \multirow{2}{*}{\textbf{Tool}} & \multicolumn{5}{c}{\textbf{ReAct Agents}} \\
\cmidrule(lr){3-7}
& & \textbf{RSH} & \textbf{GEN} & \textbf{PAT} & \textbf{CVA} & \textbf{REF} \\
\midrule
\multirow{6}{*}{\rotatebox{90}{\textit{Docs}}}
& \texttt{method\_doc} & \ding{51} & \ding{51} & \ding{51} & \ding{51} & \ding{51} \\
& \texttt{class\_doc} & \ding{51} & \ding{51} & \ding{51} & \ding{51} & \ding{51} \\
& \texttt{package\_doc} & \ding{51} & \ding{51} & \ding{51} & \ding{51} & \ding{51} \\
& \texttt{list\_packages} & \ding{51} & \ding{51} & \ding{51} & \ding{51} & \ding{51} \\
& \texttt{list\_classes} & \ding{51} & \ding{51} & \ding{51} & \ding{51} & \ding{51} \\
& \texttt{list\_methods} & \ding{51} & \ding{51} & \ding{51} & \ding{51} & \ding{51} \\
\midrule
\multirow{5}{*}{\rotatebox{90}{\textit{Code}}}
& \texttt{get\_method\_code} & \ding{51} & \ding{51} & \ding{51} & \ding{51} & \ding{51} \\
& \texttt{get\_class\_code} & \ding{51} & \ding{51} & \ding{51} & \ding{51} & \ding{51} \\
& \texttt{find\_definition} & \ding{51} & \ding{51} & \ding{51} & \ding{51} & \ding{51} \\
& \texttt{find\_refs} & \ding{51} & \ding{51} & \ding{51} & \ding{51} & \ding{51} \\
& \texttt{grep} & \ding{51} & \ding{51} & \ding{51} & \ding{51} & \ding{51} \\
& \texttt{find\_symbol} & \ding{51} & \ding{51} & \ding{51} & \ding{51} & \ding{51} \\
\midrule
\multirow{2}{*}{\rotatebox{90}{\textit{CG}}}
& \texttt{reach\_methods} & -- & -- & -- & (\ding{51}) & -- \\
& \texttt{path\_to\_method} & -- & -- & -- & \ding{51} & -- \\
\midrule
\multirow{2}{*}{\rotatebox{90}{\textit{Exec}}}
& \texttt{compiler} & -- & -- & (\ding{51}) & -- & -- \\
& \texttt{jazzer} & -- & -- & -- & (\ding{51}) & -- \\
\bottomrule
\end{tabular}
\smallskip
\begin{flushleft}
\footnotesize
\textbf{Agents:} RSH (Research), GEN (Generation), PAT (Patching), CVA (Coverage Analysis), REF (Refinement).
\ding{51}~= Available, (\ding{51})~= Static invocation (not queryable), --~= No access.
\textbf{Categories:} Docs (Javadoc API documentation); Code (source code indexing); CG (call graph analysis); Exec (compiler and fuzzer).
\end{flushleft}
\vspace{-3mm}%
\label{tab:tool-availability}
\end{table}
\subsection{Target Research}%
\label{subsec:target-research}
Following environment initialization (Maven download, documentation extraction, callgraph construction), the research agent transforms the target method signature into contextual knowledge about API semantics. The agent is initialized with the target method's signature, documentation, and source code. Then, the agent can iteratively query additional documentation and source code through the provided tools. Figure~\ref{fig:research-sequence} illustrates this query-driven exploration pattern. Rather than exhaustively extracting all available information, the agent follows its reasoning to identify relevant patterns: required initialization sequences, factory method usage, and implicit preconditions. The agent produces a structured, natural language research report with predefined markdown sections that organize findings without constraining content to rigid schemas, accommodating diverse API designs and model outputs.

\begin{table*}[t]
\caption{Benchmark library characteristics. All libraries are widely-deployed in the Maven ecosystem and represent real-world fuzzing targets.}
\centering
\setlength{\tabcolsep}{3pt}
\renewcommand{\arraystretch}{1.1}
\small
\begin{tabular}{l r r l l l r}
\toprule
\textbf{Library} & \textbf{Version} & \textbf{Dependents} & \textbf{Class Name} & \textbf{Target Method} & \textbf{Category} & \textbf{Rank} \\
\midrule
commons-cli      & 1.10.0     & 5K+  & DefaultParser        & parse(Options, String[])              & CLI Parser       & \#1 \\
gson             & 2.13.1     & 27K+ & JsonParser           & parseString(String)                   & JSON Library     & \#2 \\
guava            & 33.4.8-jre & 42K+ & HostAndPort          & fromString(String)                    & Core Utilities   & \#1 \\
jackson-databind & 2.20.0     & 36K+ & ObjectMapper         & readTree(String)                      & JSON Library     & \#1 \\
jsoup            & 1.21.1     & 4K+  & Jsoup                & parse(String)                         & HTML Parser      & \#1 \\
\midrule
\multirow{2}{*}{antlr4} & \multirow{2}{*}{4.13.2} & \multirow{2}{*}{1K+} & Grammar & Grammar(String) & \multirow{2}{*}{Parser Generator} & \multirow{2}{*}{\#1} \\[-.9pt]
                &            &      & Grammar              & createParserInterpreter(TokenStream)  &                  &      \\
\bottomrule
\end{tabular}
\smallskip
\vspace{-3mm}%
\label{tab:benchmarks}
\end{table*}

\subsection{Harness Generation}%
\label{subsec:harness-generation}
The research report is transformed into compilable code through two sequential steps: generation and compilation. The generation agent synthesizes initial harness code that instantiates the target method with fuzzer-generated inputs. The agent has access to the Jazzer API documentation and queries additional source code to resolve ambiguities in constructor signatures or factory method usage. A critical aspect of harness synthesis is exception handling: the agent must determine which exceptions represent expected API behavior (e.g., \textit{IllegalArgumentException} for invalid inputs) that should be caught to continue fuzzing, versus unexpected exceptions that indicate bugs and must propagate to Jazzer's crash detection. The agent analyzes API documentation and method signatures to infer expected exception contracts, synthesizing appropriate try-catch blocks that preserve bug-finding capability. The agent outputs harness source code and a list of Maven dependencies. Separating research from generation prevents context saturation: research explores broadly without committing to code structure, while generation focuses narrowly on producing syntactically valid harness code.

If the compile step fails, a compilation agent iteratively resolves build errors by analyzing compiler diagnostics, querying source code and documentation to understand the root cause, and producing corrected code until compilation succeeds or an iteration limit is reached. Common error patterns include missing imports, incorrect method signatures, and improper exception handling. 

\subsection{Coverage-Guided Refinement}%
\label{subsec:coverage-guided-refinement}

Once compilation succeeds, the compiled harness is instrumented and executed under fuzzing to collect initial coverage data. An iterative refinement loop then uses this coverage feedback to improve harness effectiveness through two collaborative agents: a coverage analysis agent that interprets coverage gaps and decides whether refinement is worthwhile, and a refinement agent that modifies the harness.

To seed the coverage analysis, we merge method-level coverage data with the static callgraph to produce an annotated view showing coverage status for each reachable method, grouped by call depth from the target. The coverage analysis agent explores uncovered or partially covered methods by querying their source code and documentation to determine whether gaps reflect addressable harness deficiencies (missing input diversity, unexplored API paths) or fundamental limitations (unreachable defensive code, external I/O dependencies). The agent then makes a termination decision: stop if further refinement yields diminishing returns, or continue with a strategy targeting specific uncovered methods.

If refinement continues, the refinement agent receives the current harness code, the coverage analysis strategy (priority methods and improvement rationale), and annotated coverage data. The agent modifies the harness to exercise uncovered code paths through strategies such as diversifying input generation, invoking alternative API paths, or triggering exception handlers through edge-case inputs. The refined harness re-enters compilation and fuzzing, creating a feedback loop that continues until the coverage agent determines that the refinement yields diminishing returns or an iteration limit is reached. Convergence detection through code hashing prevents oscillation between semantically equivalent harness variants.

\section{Evaluation}%
\label{sec:evaluation}

We evaluate our harness generation approach on widely used Maven libraries, examining achieved coverage, computational costs, and agent behavior patterns. Our evaluation demonstrates that the approach produces competitive harnesses to existing baselines and techniques while maintaining practical generation costs. Additionally, during the 12-hour fuzzing campaigns with the generated harnesses, we uncovered multiple previously unknown bugs in mature libraries, validating their effectiveness in real-world scenarios.

\subsection{Experimental Setup}%
\label{subsec:exp-setup}

\begin{figure*}[t]
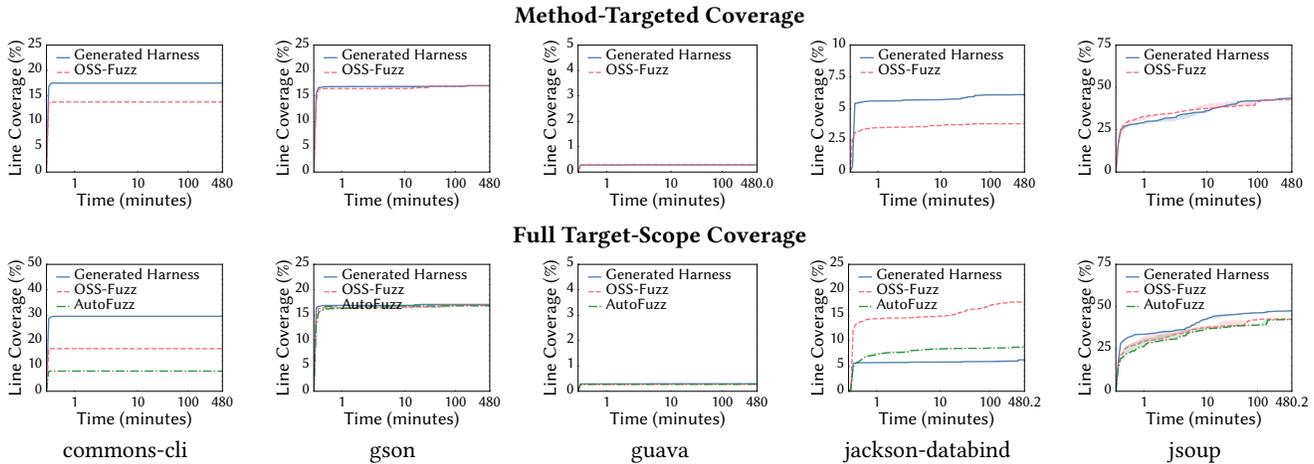

    \centering
    \textbf{Method-Targeted Coverage}\\[4pt]
    \begin{tabular}{ccccc}
        \includesvg[width=0.18\textwidth]{data/coverage/commons_coverage_comparison.svg} &
        \includesvg[width=0.18\textwidth]{data/coverage/gson_coverage_comparison.svg} &
        \includesvg[width=0.18\textwidth]{data/coverage/guava_coverage_comparison.svg} &
        \includesvg[width=0.18\textwidth]{data/coverage/jackson_coverage_comparison.svg} &
        \includesvg[width=0.18\textwidth]{data/coverage/jsoup_coverage_comparison.svg}
    \end{tabular}

    \vspace{2pt}

    \textbf{Full Target-Scope Coverage}\\[4pt]
    \begin{tabular}{ccccc}
        \includesvg[width=0.18\textwidth]{data/coverage/package_commons-cli_coverage.svg} &
        \includesvg[width=0.18\textwidth]{data/coverage/package_gson_coverage.svg} &
        \includesvg[width=0.18\textwidth]{data/coverage/package_guava_coverage.svg} &
        \includesvg[width=0.18\textwidth]{data/coverage/package_jackson-databind_coverage.svg} &
        \includesvg[width=0.18\textwidth]{data/coverage/package_jsoup_coverage.svg} \\
        commons-cli & gson & guava & jackson-databind & jsoup
    \end{tabular}

    \caption{Coverage comparison across five Java libraries with three runs over 8-hour fuzzing campaigns. \textbf{Top row:} Method-targeted coverage for our generated harnesses, focusing exclusively on target method execution. \textbf{Bottom row:} Full target-scope coverage enabling fair comparison with AutoFuzz baseline. Each plot shows average branch coverage percentage over time with min and max coverage shown as shaded regions.}
    \Description{Coverage comparison across five Java libraries with three runs over 8-hour fuzzing campaigns. \textbf{Top row:} Method-targeted coverage for our generated harnesses, focusing exclusively on target method execution. \textbf{Bottom row:} Full target-scope coverage enabling fair comparison with AutoFuzz baseline. Each plot shows average branch coverage percentage over time with min and max coverage shown as shaded regions. In most cases the generated harnesses outperform both the OSS-Fuzz and AutoFuzz baselines.}%
    \label{fig:coverage-comparison}
\end{figure*}
We evaluate on seven target methods from six widely-deployed Java libraries (Table~\ref{tab:benchmarks}). The selected targets span parsers (commons-cli, jsoup, antlr4), JSON libraries (gson, jackson-databind), and core utilities (guava).
We compare our generated harnesses against two baselines. OSS-Fuzz~\cite{oss-fuzz} is Google's continuous fuzzing service for open source software. All selected target methods have existing harnesses in OSS-Fuzz. These harnesses serve as our primary baseline.
We additionally compare against Jazzer AutoFuzz~\cite{jazzer}, an automated harness generation mode built into the Jazzer coverage-guided fuzzer for the JVM. AutoFuzz leverages Java reflection to automatically generate harnesses by discovering accessible constructors and methods, recursively building required objects through structure-aware type instantiation. Unlike our approach, AutoFuzz operates without program analysis or coverage feedback, relying solely on runtime reflection to explore the API surface.
For LLM-based comparison, we attempted to use OSS-Fuzz-Gen~\cite{oss-fuzz-gen}, but encountered implementation issues, primarily in model output parsing, that prevented successful harness generation for our Java targets despite several attempts at fixing the underlying issues.
\par
We implement our approach using LangGraph~\cite{langgraph} for workflow orchestration and Claude 4.5 Sonnet (2025-09-29) as the underlying model. Harnesses are compiled using Gradle and executed using Jazzer with instrumented coverage collection.
\par
To measure the effectiveness of fuzzing the targeted method, we measure coverage under two configurations: (1)~method-targeted coverage activates only during target method execution (Section~\ref{subsec:static-analysis}), focusing metrics on target behavior; (2)~full target-scope coverage uses standard JaCoCo instrumentation across the entire library for a fair baseline comparison.
All campaigns are run for 12 hours per target with a single fuzzing thread and an empty seed corpus.

\subsection{Coverage Effectiveness}%
\label{subsec:coverage-effectiveness}
\begin{figure}[hb]
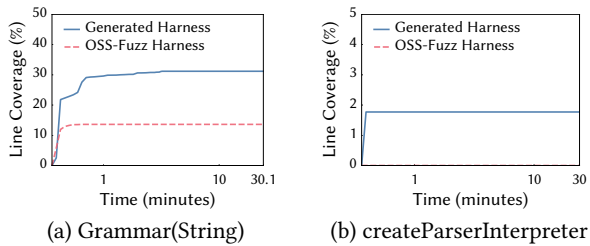

    \centering
    \begin{tabular}{cc}
        \includesvg[width=0.45\columnwidth]{data/coverage/antlr4-grammar_coverage_comparison.svg} &
        \includesvg[width=0.45\columnwidth]{data/coverage/antlr4-parser_coverage_comparison.svg} \\
        (a) Grammar(String) & (b) createParserInterpreter
    \end{tabular}
    \caption{Method-targeted coverage for ANTLR4's two target methods.}
    \Description{Method-targeted coverage for ANTLR4's two target methods, comparing OSS-Fuzz harness variants against our generated harnesses.}%
    \label{fig:antlr4-coverage}
\end{figure}

Figure~\ref{fig:coverage-comparison} shows line coverage over time across five Java libraries. The top row compares method-targeted coverage for our generated harnesses against the OSS-Fuzz baseline, evaluating focused execution of target method logic. The bottom row shows full package-scope coverage allowing comparison against AutoFuzz.
Under method-targeted coverage the generated harnesses have a median improvement of $26$\% over the OSS-Fuzz harnesses, demonstrating the effectiveness of our system in generating harnesses for specific methods. The temporal dynamics reveal that the primary difference is observable early into the fuzzing campaign, suggesting that the harness provides better structural input diversity.
Comparing under full target-scope coverage, our generated harnesses outperform the AutoFuzz and OSS-Fuzz baseline by a median of $5$\% and $6$\% respectively. The only target the generated harness does not outperform the baselines on is jackson-databind, where the  OSS-Fuzz harness contains additional fuzzing logic after the execution of the target method that causes an increase in the overall coverage.
\par
To demonstrate the benefits of targeted harness generation, we evaluate ANTLR4, which has two target methods in our benchmark. The existing OSS-Fuzz harness~\footnote{https://github.com/google/oss-fuzz/blob/master/projects/antlr4-java/GrammarFuzzer.java} exercises both methods sequentially. It creates a Grammar object from fuzzed input, then invokes \textit{createParserInterpreter} on the resulting grammar. This sequential dependency means the parser interpreter is only reached when grammar creation succeeds without throwing exceptions.
We compare our automatically generated harnesses (each targeting one method individually) against the OSS-Fuzz harness measured with method-targeted coverage scoped to each target method. For the Grammar constructor (Figure~\ref{fig:antlr4-coverage}(a)), we evaluate both the unmodified OSS-Fuzz harness and a manually edited variant that only creates the grammar without invoking the parser interpreter. For \textit{createParserInterpreter} (Figure~\ref{fig:antlr4-coverage}(b)), we evaluate only the unmodified OSS-Fuzz harness.
Our generated harnesses outperform the OSS-Fuzz baseline in both scenarios. Most notably, the OSS-Fuzz harness achieves 0\% coverage for \textit{createParserInterpreter} throughout the campaign, indicating that grammar creation consistently throws exceptions before reaching the parser interpreter call. In contrast, our generated harness successfully exercises this method by synthesizing inputs that satisfy the grammar constructor's preconditions. Note that both ANTLR4 campaigns encountered a Jazzer timeout at 30 minutes that terminates Jazzer execution. However, the coverage trends before termination demonstrate the performance difference.
\subsection{Bug Discovery}%
\label{subsec:bug-discovery}
Beyond coverage metrics, we examine whether our generated harnesses discovers novel bugs during fuzzing campaigns. Jazzer reports crashes through its exception handling infrastructure, distinguishing between genuine crashes (uncaught exceptions indicating bugs) and caught exceptions that represent normal control flow. This represents a core challenge for the harness generation, as the generated harnesses must avoid over-catching exceptions that would mask real bugs while also separating spurious crashes from expected error handling.
Across a single 12-hour fuzzing campaign, the generated harnesses triggered a total of 14 crashes in two libraries. After manual investigation, we determined that all reported crashes represent genuine bugs in the target library. The harnesses correctly identified uncaught exceptions rather than reporting false positives from expected exception handling. Manual triage revealed 3 unique bugs:

\begin{itemize}
\item \textbf{commons-cli}: Two distinct null pointer exceptions in option parsing logic, triggered by edge-case combinations of option configurations and malformed arguments. The crashes manifest in both long-option and short-option code paths, with 12 total crash artifacts reducing to 2 unique root causes.
\item \textbf{jsoup}: One index-out-of-bounds exception in HTML tree building logic, triggered by complex malformed HTML input ($\sim 1KB$). This crash represents a potential denial-of-service vector, as attackers could craft HTML to crash the parser. Two crash artifacts correspond to the same underlying bug.
\end{itemize}

These results demonstrate that our automatically-generated harnesses achieve sufficient input diversity and API coverage to discover real bugs in mature, widely-deployed libraries that already have existing harnesses in the OSS-Fuzz ecosystem. The fact that Jazzer reported zero false positives highlights the effectiveness of the harness generation in handling exceptions precisely to expose genuine bugs without over-catching.
\begin{table}[htb]
\caption{Harness generation cost and agent activity across target libraries. Agent rows show iterations and tool calls, with multiple values indicating successive refinement rounds.}
\centering
\setlength{\tabcolsep}{3pt}
\renewcommand{\arraystretch}{1.1}
\small
\begin{tabular}{l l | r r r r r | r}
\toprule
& \textbf{Metric} & \textbf{commons-cli} & \textbf{gson} & \textbf{guava} & \textbf{jackson} & \textbf{jsoup} & \textbf{Avg} \\
\midrule
\multirow{3}{*}{\rotatebox{90}{\textbf{Total}}} & Tokens & 1896K & 557K & 395K & 1285K & 1039K & 982K \\
& Cost & \$6.25 & \$1.81 & \$1.34 & \$4.13 & \$3.32 & \$3.20 \\
& Time & 1200s & 337s & 417s & 684s & 551s & 599s \\
\midrule
\multirow{2}{*}{\rotatebox{90}{\textbf{RES}}} & Iter & 17 & 17 & 16 & 17 & 19 & 17.7 \\
& Tools & 44 & 27 & 23 & 24 & 28 & 30.7 \\
\midrule
\multirow{2}{*}{\rotatebox{90}{\textbf{GEN}}} & Iter & 3 & 2 & 3 & 3 & 4 & 3.1 \\
& Tools & 5 & 2 & 2 & 3 & 4 & 3.4 \\
\midrule
\multirow{2}{*}{\rotatebox{90}{\textbf{PAT}}} & Iter & 5/0/0/0 & 0 & 0/0 & 6/0 & 0/0 & 0.9 \\
& Tools & 11/0/0/0 & 0 & 0/0 & 6/0 & 0/0 & 1.2 \\
\midrule
\multirow{2}{*}{\rotatebox{90}{\textbf{CVA}}} & Iter & 16/14/13/14 & 17 & 4/4 & 18/14 & 14/18 & 12.5 \\
& Tools & 28/22/16/28 & 31 & 4/5 & 21/16 & 24/21 & 18.3 \\
\midrule
\multirow{2}{*}{\rotatebox{90}{\textbf{REF}}} & Iter & 10/7/15 & -- & 5 & 8 & 10 & 8.9 \\
& Tools & 19/15/23 & -- & 5 & 13 & 22 & 16.1 \\
\bottomrule
\end{tabular}
\smallskip
\begin{flushleft}
\footnotesize
\textbf{Agents:} RES (Research), GEN (Generation), PAT (Patching), CVA (Coverage Analysis), REF (Refinement). Iter = ReAct Iterations, Tools = Tool Calls.
\end{flushleft}
\vspace{-3mm}
\label{tab:generation-cost}
\end{table}

\subsection{Generation Costs and Agent Behavior}%
\label{subsec:generation-costs}
Table~\ref{tab:generation-cost} details the computational costs and agent activity patterns across the five main targets. Harness generation costs range from \$1.34 (guava) to \$6.25 (commons-cli), with an average of \$3.20 per harness. Generation completes in an average of 599 seconds, making the approach practical for integration into iterative fuzzing workflows.
Token consumption directly correlates with workflow complexity. The observed diversity in workflow iterations and agent loops indicates effective adaptation to target-specific characteristics, with more complex APIs requiring additional research and refinement cycles while simpler targets lead to quicker convergence. 
\par
Comparing usage patterns between the research agent and the generation agent reveals that the initial report contains most of the necessary information for synthesis requiring on average only $3.1$ iterations until the first harness is synthesized. Additionally, the patching agent invocation pattern reveals that, in most cases, the initial synthesis is already correct and only few targets require repair iterations with all evaluated harnesses compiling successfully after at most $6$ iterations.
The coverage analysis and refinement agents show high variance reflecting target-specific characteristics.
This adaptive behavior demonstrates that our agent-based termination successfully distinguishes between targets where refinement yields benefits and those where additional iteration would waste resources.

\section{Related Work}
\label{sec:related-work}

\paragraph{Classical Harness Generation}
Different traditions of program analysis have shaped how fuzzing harnesses are constructed. \textit{Usage-based generation} mines valid API calls from existing consumer code or unit tests, as demonstrated by systems that slice client code into reusable API snippets~\cite{DBLP:conf/sigsoft/BabicBCIKKLSW19:FUDGE}, construct harness stubs from API dependence graphs~\cite{DBLP:conf/uss/IspoglouAMP20:FuzzGen}, or inject fuzzed inputs into test cases~\cite{DBLP:conf/sp/JeongJYMKJKSH23:UTopia}. While this captures realistic interaction patterns, it remains dependent on the availability of suitable consumer code. In contrast, \textit{structure-based generation} derives harnesses directly from type signatures and interface specifications, building dataflow graphs to capture API interactions~\cite{DBLP:conf/icse/GreenA22:GraphFuzz} or introducing intermediate representations for large-scale libraries~\cite{DBLP:journals/pacmse/ToffaliniBTP25:LibErator, DBLP:conf/icse/ShermanN25:OGHarn}. These approaches offer broader applicability but often lack iterative refinement, leaving adaptation to new targets largely manual. \textit{Feedback-driven generation} refines harnesses iteratively using runtime signals, employing automaton learning on API usage patterns~\cite{DBLP:conf/uss/ZhangLZZZZXLL0H23:Rubick} or validating candidates using compile-time and runtime oracles~\cite{DBLP:conf/icse/ShermanN25:OGHarn}. While promising, such systems frequently rely on domain-specific heuristics or fixed coverage thresholds for termination decisions.

\paragraph{LLM-based Harness Synthesis}
With the availability of large language models, harness generation has been explored from a learning perspective. Early feasibility studies evaluate prompting strategies~\cite{DBLP:conf/issta/ZhangZBLMXLSL24:HowEffectiveAreThey} and identify obstacles such as semantic drift~\cite{DBLP:conf/sigsoft/Jiang0MCZSWFWLZ24:WhenFuzzingMeetsLLMs}. Recent systems demonstrate automatic synthesis through coverage-guided prompt mutation for iterative refinement~\cite{DBLP:conf/ccs/LyuXCC24:PromptFuzz}, which mutates prompts based on coverage feedback but lacks semantic interpretation of coverage gaps. CKGFuzzer~\cite{DBLP:conf/icse/XuMZZCHLW25:CKGFuzzer} augments LLM reasoning with knowledge graphs of API relations, preprocessing entire API surfaces into static graphs before generation. Other approaches integrate LLM reasoning into static analysis pipelines~\cite{DBLP:journals/corr/abs-2505-03425:HGFuzzer} or combine LLM-based repair with solver-driven scheduling~\cite{DBLP:journals/corr/abs-2507-18289:Scheduzz}. However, most approaches either preprocess all information upfront (risking context saturation and prioritizing breadth over target-specific depth) or lack mechanisms for agents to iteratively query documentation and source code as reasoning progresses. Our work provides agents with query-based access to documentation, source code, and callgraph information through the Model Context Protocol~\cite{mcp}, enabling on-demand retrieval as reasoning needs emerge. While practical tools like OSS-Fuzz-Gen~\cite{oss-fuzz-gen} employ comparable multi-agent workflows with generic source access and fixed iteration counts, we provide specialized static analysis (callgraph construction, method-targeted coverage) and adaptive orchestration where coverage analysis agents determine iteration budgets dynamically. We extend feedback-driven refinement~\cite{DBLP:conf/ccs/LyuXCC24:PromptFuzz, DBLP:conf/uss/ZhangLZZZZXLL0H23:Rubick} by delegating termination decisions to agents that interpret coverage gaps, rather than applying fixed thresholds or heuristics.

\section{Conclusion}

We presented a multi-agent architecture that automates fuzzing harness generation for Java libraries through specialized ReAct agents and query-driven code analysis. Five agents decompose the workflow into research, synthesis, compilation repair, coverage analysis, and refinement, querying documentation and source code on demand through the Model Context Protocol. Method-targeted coverage instrumentation and agent-guided termination enable effective refinement without context saturation.

Evaluation on seven target methods from six widely deployed libraries demonstrates competitive coverage with OSS-Fuzz baselines at practical costs of \$3.20 and 10 minutes per harness. Our harnesses discovered 3 bugs in production libraries already integrated into OSS-Fuzz, validating that automated generation achieves sufficient quality for real vulnerability discovery. Future work includes extending to the system stateful APIs, identifying minimal method sets that maximize library coverage to reduce generation costs, and adapting the approach to synthesize property-based security harnesses that verify invariants beyond crash detection.

\begin{acks}
This work has been supported by funding from the Agentur für Innovation in der Cybersicherheit GmbH (Cyberagentur, project SOVEREIGN)
\end{acks}
\bibliographystyle{ACM-Reference-Format}
\bibliography{main}

\end{document}